# Transforming Discarded Thermoelectrics into High-Performance HER Catalysts

Gemeda Jemal Usa[a†], Caique C. Oliveira[b†], Varinder Pal[c†], Suman Sarkar[d], Gebisa Bekele Feyisa[a*], Moumita Kotal[c*], Emmanuel Femi olu[e*], Pedro A. S. Autreto[b*], Temesgen Debelo Desissa[a*], and Chandra Sekhar Tiwary[c*]

[a]Materials Science and Engineering, College of Mechanical, Chemical and Materials Engineering, Adama Science and Technology University, Adama, 1888, Ethiopia

[b]Center for Natural and Human Sciences (CCNH) - Federal University of ABC (UFABC), Santo André - Brazil

[c]Department of Metallurgical and Materials Engineering, Indian Institute of Technology, Kharagpur, 721302, West Bengal, India

[d]Materials Engineering Department, Indian Institute of Technology, Jammu, 182121, Jammu and Kashmir, India

[e]Faculty of Materials Science and Engineering, Jimma Institute of Technology, Jimma University, P.O. Box 378, Jimma, Oromia, Ethiopia

*gebisabek12@gmail.com

*moumita.kotal@gmail.com

*foluemm@gmail.com

*pedroautreto@gmail.com

*temesgen.debelo@astu.edu.et

*chandra.tiwary@metal.iitkgp.ac.in

† Equal Contribution

**Abstract**

With the increase in the complexity of the materials used in various sophisticated electronic devices, recycling of E-waste is more challenging. In the present study, we have converted thermoelectric (TE) waste into functional HER electrocatalyst by considering circular-economy and low-carbon approach. The as received TE waste was processed through ball milling (TE waste-BM) and melting casting (TE waste-M) routes. Morphological and structural evaluations revealed that the formation of $BiSbTe_3$/ZnTe heterostructure in TE-waste-M promote HE efficiency when compared to the presence of $Bi_2Te_3$/$BiSbTe_3$ heterostructure (TE-waste-BM). TE waste-M exhibited lower overpotential (641 mV at 10 mA cm$^{-2}$), smaller Tafel slope (233 mV dec$^{-1}$) and stable operation for 5.5 h with negligible current decay than that of TE waste-BM, attributed to the accelerated charge transfer, fast water dissociation steps and rapid hydrogen adsorption in TE waste-M, originated from the presence of $BiSbTe_3$/ZnTe heterostructure, defect enriched interfaces. Density functional theory calculations supported the experimental findings, revealing that heterostructures strengthens the bonding states near the Fermi level, thereby enhancing the HER activity of $BiSbTe_3$/ZnTe heterostructure. This work simultaneously integrates waste management with green hydrogen production by offering an economically viable, scalable and low-carbon approach for HER catalysts.

**Keywords:** Thermoelectric waste, Hydrogen evolution, density functional theory

## 1. Introduction

The rapid decline of fossil fuel resources, coupled with escalating climate challenges, has necessitated the urgency for sustainable and carbon-neutral energy solutions.[1] Among the available options, hydrogen stands out as a clean fuel with exceptionally high gravimetric energy density, positioning it as a pivotal element in the transition toward a green economy. Electrochemical water splitting is regarded as one of the most promising approaches for large-scale hydrogen production, as it enables the direct conversion of renewable electricity into chemical fuel by decomposing water into hydrogen and oxygen.[2] The hydrogen evolution reaction (HER), a key half-reaction in water splitting, plays a critical role in determining the efficiency and practicality of sustainable hydrogen production.[3] Particularly, HER in alkaline electrolyte exhibits several distinct benefits- long-lasting operational stability and suitability with abundant catalysts, governing it as a preferable approach for large-scale hydrogen production.[4,5] However, it suffers from sluggish kinetics in alkaline media due to the additional energy barrier associated with water dissociation compared to that in acidic media, thereby urgently necessitating the development of highly efficient, cost-effective and durable catalysts.

The critical requirements of conventional HER catalysts involve refinement of metallic precursors, numerous chemical treatment steps, and energy-consumptive mineral extraction, all of which lead to a higher carbon footprint.[6,7] Also, the extraction and refinement of critical metals create major environmental impact, like toxic sewages, soil contamination and greenhouse gas emissions.[8] To mitigate all these global issues, the development of efficient and environmentally sustainable electrocatalyst for green hydrogen production is highly desirable.

Motivating from circular economy approach and low-carbon electrochemical technologies, valorization of solid wastes as catalytic precursors is an attractive technique for HER, which can bridge the green energy conversion with waste product valuation.[9] In particular, biowaste-derived carbon materials generally depict inadequate catalytic activity for alkaline HER owing to their low electrical conductivity and less available active sites and therefore require additional functionalization. Incorporation of transition metals (e.g., Co, Ni, Mo) or heteroatom doping (e.g., N, S, P) is typically necessary to create active sites, enhance conductivity, and improve overall HER performance.[10,11] In view of these discrepancies, thermoelectric waste (TEW) from discarded modules and electronic devices offers unique advantages for alkaline HER owing to their tunable electronic structures, multi-component active sites, intrinsic defect, inherent heterojunctions and copious grain boundaries. The components of TEW are typically rich in transition metals, chalcogenides, and dopants (e.g., Bi, Sb, Zn, Te, Se, Ni, Co), which can be transformed into catalytically active phases through electrolysis. Repurposing TEW not only diverts hazardous electronic materials from landfills but also eliminates the need for additional mining or precursor synthesis, significantly lowering lifecycle energy consumption and associated emissions. Moreover, the essential features including intrinsic structural heterogeneity, defect-rich surfaces, and favorable electronic configurations stemming from their original thermoelectric function, are beneficial for accelerating charge transfer and hydrogen adsorption during HER. As a result, TEW can promote water dissociation (Volmer step) expediting charge transfer and hydrogen production without employing any complex synthetic steps.

Motivating by these advantageous features of TEW, the present work deals with the valorization of TEW into functional electrocatalyst representing a recycling management, which not only mitigate the environmental impact and carbon footprint but also reduce the reliance on less abundant resources and offer a scalable, cost-effective approach for alkaline HER. TE waste was processed using ball milling and melting casting approaches. These processed materials were characterized at various stages to understand morphological and structural changes using X-ray diffraction (XRD), Scanning electron microscopy (SEM), and X-ray photoelectron microscopy (XPS) analyses. At last, the processed materials were then used as HER electrocatalysts and investigated the comparative effect of structural and electronic contribution of the two designed TEWs towards efficient sustainable hydrogen production in alkaline electrolyte. First principles calculations based on Density Functional Theory were carried out to investigate the activity of different structures based on experimental results, strengthening the understanding of HER behavior. This sustainable route resonates with global efforts to design circular, carbon-neutral hydrogen technologies and opens a new avenue for valorization of electronic waste to functional catalytic materials towards scalable green hydrogen production.

## 2. Experimental and Simulation Details

### 2.1 Experimental Details:

**Figure 1** shows the top view of an opened commercial TE module. As seen, **Fig. 1**, shows the TE legs of a used device, and these legs were precisely removed from the module. After removing the legs, some of these legs were processed using vibratory ball mill and the powder thus obtained was termed as TE waste-BM. Out of 5 gm of legs that removed from the TEW were taken, 2.5 gm of legs were ball milled using a vibratory ball mill for 2 hours to get a homogenous powder. The remaining 2.5 gm was melted first by using a flame, then by tungsten inert gas arc to produce a homogenous melted alloy. The ingot thus obtained after the melting was termed as TE waste-M. All the melting processes were performed in an argon atmosphere. Thus, different alloys investigated in the present study are named TE waste, TE waste-BM and TE waste-M for as received TE legs, ball milled TE waste, and melted TE waste respectively.

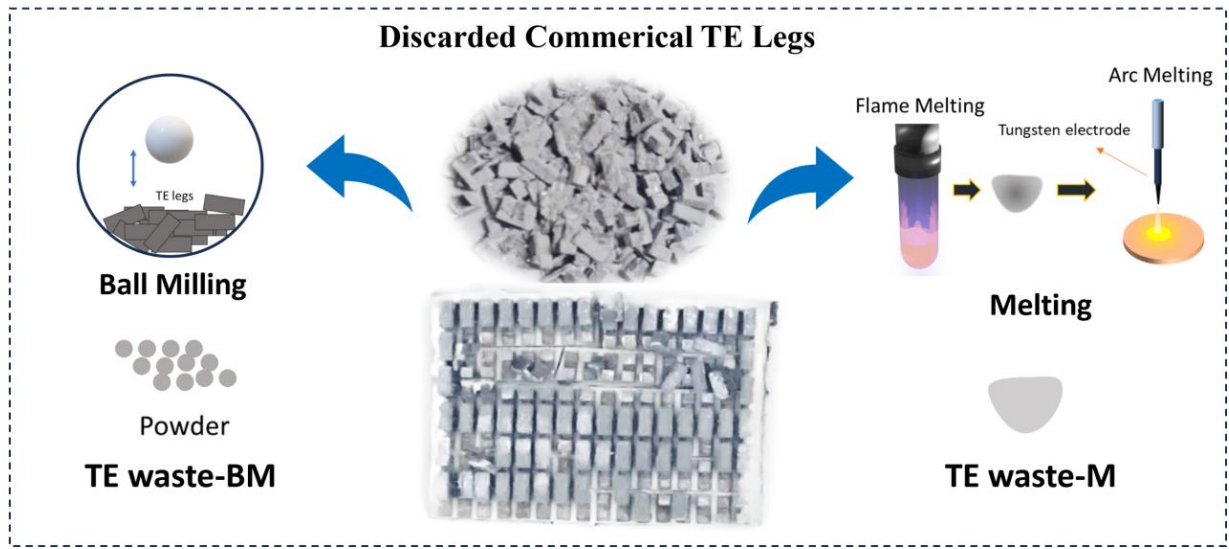

**Fig. 1** Used thermoelectric device, TE legs separated from device, schematic illustrating the process of powder preparation and melting of separated TE legs.

X-ray diffraction (Bruker D8 ADVANCE powder diffractometer) was used to identify the phases in as cast, ball milled and melted alloys. In addition, scanning electron microscope (SEM, JEOL-IT300HR (LA)) was used to understand the developed microstructure, alloy composition before and after the hydrogen evolution reaction (HER). XPS measurements (PHI 5000 VERSA PROBE III) were also performed to understand the changes in chemical bonding of different alloys before and after HER. HER measurements were performed in 1M KOH at 298 K by using a three-electrode system.

## 2.2. Computational Methods

First principles calculations were carried out within the Density Functional Theory (DFT) as implemented in the *Vienna ab initio Simulation Package* (VASP) [12] package with the Generalized Gradient Approximation for the exchange and correlation functional with dispersion corrections with Becke-Johnson damping function (DFT-D3 functional).[13] Core electrons were approximated with pseudopotentials within the projected augmented wave (PAW) method as implemented in

VASP.[14] Self-consistency threshold was set to $10^{-7}$ eV, and forces were minimized below 0.02 eV/A. Brillouin zone sampling was performed only at the Gamma point for geometry optimization calculations, while for electronic structure, (4x4x2) Monkhorst and Pack grid [15] was employed.

The catalytic activity was calculated within the Computational Hydrogen Electrode (CHE) model [16] by calculating the H adsorption free energy ($\Delta G_H$):

$$\Delta G_H = E_{H\cdot} - \left(E_\cdot + \frac{1}{2}E_{H_2}\right) + \Delta E_{ZPE} - T\Delta S$$

Where $E_{H\cdot}$ represents the total energy of the optimized slab with an adsorbed H, $E_\cdot$ the total energy of the optimized and clean slab, and $E_{H_2}$ the energy of a hydrogen molecule in gas phase. The last two terms represent the zero-point energy and entropy change with respect to the gas phase, respectively. In this work, the value of 0.24 eV was employed for these two terms, following previous works.[17] Crystal Orbital Hamilton Population (COHP) analysis was performed via LOBSTER package.[18,19] Within this scheme, a good catalyst exhibits displays $\Delta G_H \approx 0$.[19] Bader charges [19] were obtained using the Henkelman group software.[20–22]

## 3. Results and Discussion

### 3.1. Before HER characterization

XRD patterns of the samples are shown in **Fig. 2a**. ICSD number: 98-015-4490, 98-007-0006, 98-010-4196, 98-004-8324, 98-010-8419, 98-000-6648 and 98-006-2949, were used as reference for $ZnO$, $SiO_2$, $\alpha$ (ZnTe), $\beta$ ($Bi_2Te_3$), $\gamma$ ($Cu_5Zn_8$), BST ($BiSbTe_3$), and Zn respectively. XRD pattern of the TE waste shows the presence of $ZnO$, $SiO_2$, $\beta$ ($Bi_2Te_3$), BST, and $\alpha$ (ZnTe) phases. In addition to the ZnO, there are small intensity peaks matching with the Zn phase in the TE waste sample. The ball milled sample (TE waste-BM) shows the presence of BST, $\beta$ and Zn phases in the alloy. On the other hand, the melting of the TE legs shows the presence of BST, $\alpha$, $\beta$ and $\gamma$ ($Cu_5Zn_8$) phases in the melted alloy (TE waste-M). To further understand the microstructural changes in the materials, SEM equipped with EDS was used and recorded micrographs are shown in **Figs. 2b-g**. The microstructure of the as received TE waste shows the presence of a grey region (oxide) containing Zn, Mg, Si, etc. while the brighter region corresponds to the Zn, Te, Bi, Sb rich regions (shown in BSE micrograph in **Fig. 2c**). On ball milling the metallic part (the primarily the TE legs), it shows the overall elemental composition of powder was found to be Zn-12.5Sb-48.45Te-19.51Bi (at.%). In addition. **Fig. 2d** shows the variation of the particle size and morphology due to the ball milling. In **Fig. 2e**, the presence of the $Bi_2Te_3$ phase was observed in the TE waste-BM. **Figs. 2f** and **g** show the microstructure of the TE waste-M. It was observed that the large and brighter regions show elemental composition matching to that of BST phase, while the greyish region corresponds to the ZnTe composition.

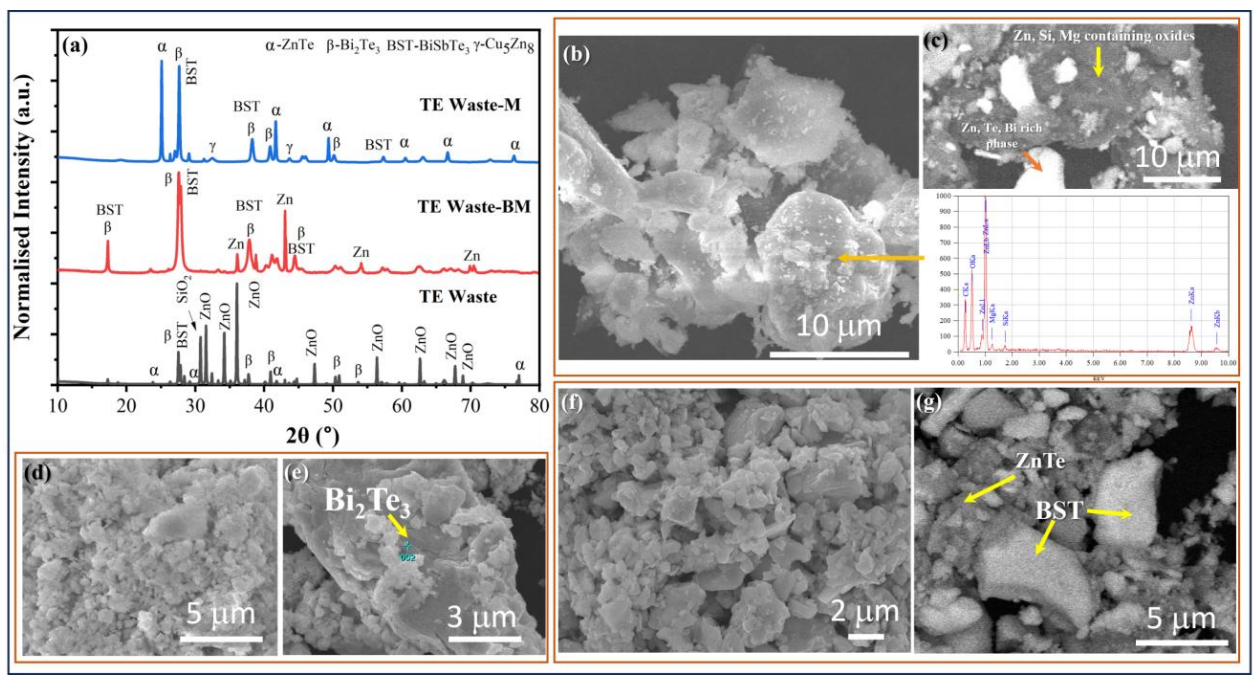

**Fig. 2** (a) XRD patterns of the investigated alloys, SEM micrograph of the (b)-(c) as received TE waste, (d)-(e) TE waste-BM, and (f)-(g) TE waste-M.

The surface chemical composition and oxidation states of the TE waste-BM were investigated by XPS (shown in **Fig. 3a-e**) confirming the presence of Te, Sb, Bi, Zn, and O elements, indicating retention of the primary constituents even after post-milling. C1s spectrum (**Fig. 3a**) displays three components-284.6 eV (C=C), 285.4 eV (C-O) and 288.2 eV (O-C=O). The prominent peak at 284.6 eV arises from adventitious carbon due to the air-exposer on the surfaces. The appearance of higher binding energy indicates surface functionalization or mild oxidation of TEW. **Figure 3b** depicts typical doublet with high intensity of Zn $2p_{3/2}$ and Zn $2p_{1/2}$ at around 1021.75 eV and 1044.76 eV corresponding to $Zn^{2+}$ in ZnO, implying that Zn remains in an oxidized environment. Also, a pair of lower-energy shoulders appeared at around 1023.22 eV and 1046.57 eV revealing the formation of $Zn(OH)_2$ or traces of Zn-Te bonding during ball milling of TE wastes, agree with XRD findings. Notably, Te3d region (**Fig. 3c**) depicts two distinct chemical states-$Te^{2-}$ (Te $3d_{5/2}$-$3d_{3/2}$ doublets at around 570.91/581.33 eV) and elemental $Te^0$ (Te $3d_{5/2}$-$3d_{3/2}$ doublets at around

574.64/585.04 eV), confirming the absence of oxidized Te, i.e, $Te^{4+}/Te^{6+}$. The Bi 4f core-level spectrum (**Fig. 3d**) exhibits two characteristic peaks at around 157.6 eV ($4f_{7/2}$) and 162.8 eV ($4f_{5/2}$), corresponding to $Bi^{3+}$ species along with weak pair of peaks of $Bi^0$ (156.11 and 161.57 eV). Therefore, Bi in the TE waste-BM exists predominantly as $Bi_2O_3$ with increase in the intensity of satellite features due to increased surface oxygen defects. **Figure 3e** presents the deconvoluted spectrum of Sb3d appearing at around 529 eV and 531.22 eV corresponding to $Sb^{3+}$ ($Sb_2O_3$) and $Sb^{5+}$ ($Sb_2O_5$), respectively, revealing the coexistence of mixed-valence antimony with partial surface oxidation. Collectively, ball milling significantly alters the surface chemistry of TEW while preserving reduced Te phases, with mild surface oxidation occurring primarily on Sb, Zn, and Bi components. Interestingly, XPS spectra of TE waste-M (**Figs. 3f-j**) reveals substantial oxidation effects induced by high-temperature processing. C 1s features arise from adventitious and oxygenated carbon (**Fig. 3f**). Zn 2p spectra (**Fig. 3g**) shows the presence of $Zn^{2+}$ features along with satellite, confirming formation of $ZnO/Zn(OH)_2$. In addition to $Te^{2-}$ and $Te^0$ components, relatively stronger $Te^{4+}$ component is found (**Fig. 3h**), revealing significant oxidation of tellurium during melting. Also, Bi4f region appears peaks corresponding to $Bi^0$ and dominant $Bi^{3+}$ component, demonstrating major oxidation upon melting (**Fig. 3i**). Similar findings have been observed for Sb3d spectrum (**Fig. 3j**) corresponding to $Sb^{3+}$ and $Sb^{5+}$, indicating of substantial oxidation of Sb on the surface. Therefore, melting shows the presence of $Te^{4+}$ and $Sb^{5+}$ along with measurable binding energy shifts in contrast to the TE waste-BM, where Te predominantly retains its reduced telluride form. Therefore, melting promotes electronic reconFig.uration and formation of different heterostructures which assist to improve charge separation by creating abundant defect sites as well as hydrogen adsorption which are crucial for enhancing HER.[23,24] Details about the role of the different heterostructures has been discussed in the section 3.3.

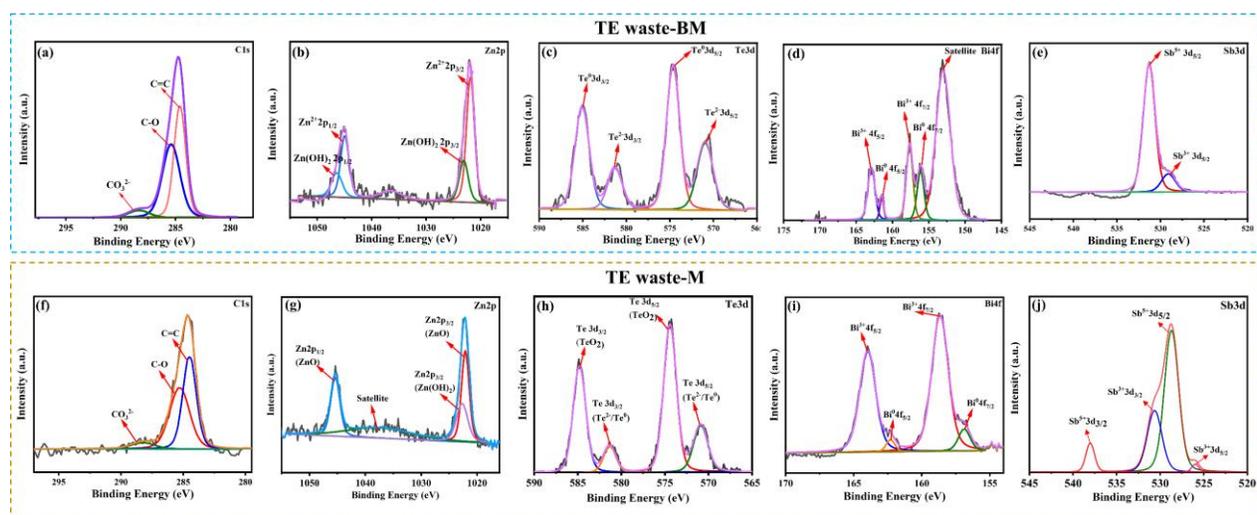

**Fig. 3** XPS spectra of (a) C 1s, (b) Zn 2p, (c) Te 3d, (d) Bi 4f, and (e) Sb 3d of TE waste-BM, (f) C1s, (g) Zn 2p, (h) Te 3d, (i) Bi 4f, and (j) Sb 3d spectra of TE waste-M.

### 3.2. Hydrogen evolution reaction

In alkaline medium, HER activity of $Bi_2Te_3$ waste-based catalysts obtained by melting and ball milling was evaluated through linear sweep voltammetry as shown in **Fig. 4a**. It is found that the TE waste-M based catalysts showed a lower onset potential (417 mV) and higher current density (~117 mA cm$^{-2}$) compared to those for ball-milled one (521 mV, 46 mA cm$^{-2}$), signifying faster kinetics for water dissociation and proton reduction. Notably, the TE waste-M catalyst exhibited a distinct smaller overpotential at 10 mA cm$^{-2}$ ($\eta_{10}$ ~ 641 mV) compared to ball-milled one (~723 mV), indicating higher efficiency for HER (**Fig. 4b**). Furthermore, the overpotential of the melted sample remained consistently lower across a wide range of current densities, as confirmed by the comparative overpotential plots with current density (**Fig. 4b**). This may be due to the possible structural differences of TE waste-M catalyst-larger crystallite size (D ~ 43 nm) with negligible lattice strain ($\varepsilon$ ~ -6.6× 10$^{-6}$) for melted one compared to those for ball milled one (D ~ 13.6 nm, $\varepsilon$ ~ 1.2× 10$^{-5}$). The presence of large, well-ordered crystallites with minimal strain promotes long-range electronic transport and stabilizes the surface electronic environment, which is essential for efficient water dissociation during the Volmer step ($H_2O + e^- \rightarrow H^* + OH^-$).[25,26] Also, the ordered crystalline domain facilitates near-optimal hydrogen adsorption free energy ($\Delta G_{H^*}^{\#}$), enabling faster kinetics for the subsequent Heyrovsky ($H^* + H_2O + e^- \rightarrow H_2 + OH^-$) and/or Tafel ($2H^* \rightarrow H_2$) steps. Moreover, the formation of α-ZnTe/β-$Bi_2Te_3$ heterojunction in the melted TE waste accelerated the interfacial charge separation and lowered charge-transfer resistance and thereby reducing overpotential. Yang et al. also found that ZnTe inclusion improved hydrogen adsorption energetics and optical absorption:-key parameters for enhanced catalytic activity.[27] Also the electronic tuning of Te-centered adsorption at the heterojunction facilitated Volmer step, i.e, water dissociation step.[28] Consequently, the HER proceeds with more facile charge transfer kinetics, reflected in a lower Tafel slope as evident from **Fig. 4c**- melt one showing significantly lower Tafel slope of 233 mV dec$^{-1}$ compared to ball milling (284 mV dec$^{-1}$). Similar findings are also observed for Ru nano catalysts which reveals that engineering well-defined crystalline domains with abundant low-coordination edge sites and oxygen vacancies at domain boundaries remarkably strengthens water adsorption and optimizes hydrogen binding at the coordinated Ru sites.[29] On the contrary, the ball-milled $Bi_2Te_3$ waste, despite its smaller crystallite size, suffers from substantial lattice strain introduced during high-energy ball milling. Such strain-induced structural deformation disrupt long-range crystallinity and create disordered grain boundaries and partially amorphous regions, impeding electron transport and thereby may destabilizing the adsorption of H*.[30] Furthermore, the TE waste-BM contains metallic Zn which rapidly passivates to $Zn(OH)_2$/ZnO in alkaline media, leading to show sluggish HER kinetics. Therefore, Volmer step is kinetically constrained followed by sluggish Heyrovsky and Tafel steps limiting the overall HER efficiency. As a result, TE waste-BM exhibited higher overpotential, lower current density and higher Tafel slope as evident from **Figs. 4a-c**. To gain deeper understanding, the electrochemically active surface area (ECSA) was estimated from the double-layer capacitance ($C_{dl}$) values derived from CV analyses in the non-faradic region at varying scan rates (**Figs. S1a-b**). Notably, the TE waste-M exhibited higher $C_{dl}$ of 0.321 mF cm$^{-2}$, while TE waste-BM showed only 0.03 mF cm$^{-2}$, indicating a larger number of accessible active sites (ECSA~ 8.03 cm$^2$) in the TE waste-M, thereby contributing to its superior HER efficiency (**Fig. 4d**). The higher efficiency is may be due to its higher crystallinity and larger crystallite size which facilitate fast charge transfer and reduce surface defects.[31] In contrast, TE waste-BM showed high microstrain and existence of less-ordered structure, which may cause agglomeration and surface reconstruction, resulting in lowering number of electrochemically available active sites (0.75 cm$^2$) and retarding charge transfer and thereby reducing HER efficiency. Lattice strain and the presence of disordered structure in the catalyst are responsible for electronic decoupling and lowering HER activity, in line with earlier reports.[32,33] To probe the charge transport dynamics, electrochemical impedance spectroscopy (EIS) was performed for both the TE waste samples as presented in **Fig. 4e**. The Nyquist plots imply that the TE waste-M possessed significantly lower semicircular diameter compared to the ball milled one, suggesting of a lower charge transfer resistance ($R_{ct}$ ~150 ohm). The reduced $R_{ct}$ indicates more efficient electron transport at the electrode-electrolyte interface, consistent with the higher crystallinity and larger crystallite size, which promote delocalized electronic pathways. On contrary, the TE waste-BM exhibited a significantly larger semicircle, revealing of sluggish charge-transfer kinetics, owing to the pronounced microstrain which is found to suppress surface reconstruction and available active sites, also thereby reducing HER efficiency. Notably, melting path is found to accelerate electronic reconfiguration for the formation of heterojunctions, as evident from XPS results, which is beneficial for faster hydrogen adsorption. To further evaluate the stability of TE Waste-BM and TE Waste-M, chronoamperometry and potential cycling method have been performed for 5.5 h as depicted in **Fig. 4f** and its inset. Both catalysts exhibit an initial fast rise in current within 40-45 minutes, which is attributed to surface restructuring, formation of steady electrode-electrolyte interface. Notably, TE waste-M shows higher current density of around -10 mA cm$^{-2}$ compared to TE waste-BM (-3 mA cm$^{-2}$) after 5 hrs, suggesting significant catalytic activity for HER. Also, TE waste-M exhibits least current decay through the entire 5 hrs, while TE waste-BM shows rapid current decay, implying sustained operational stability in alkaline medium, further reaffirming its structural robustness even after continuous operation. Similar findings are also observed for the corresponding LSV plot after 220 cycles with insignificant loss of current response, revealing promising durability of TE waste-M as HER catalyst.

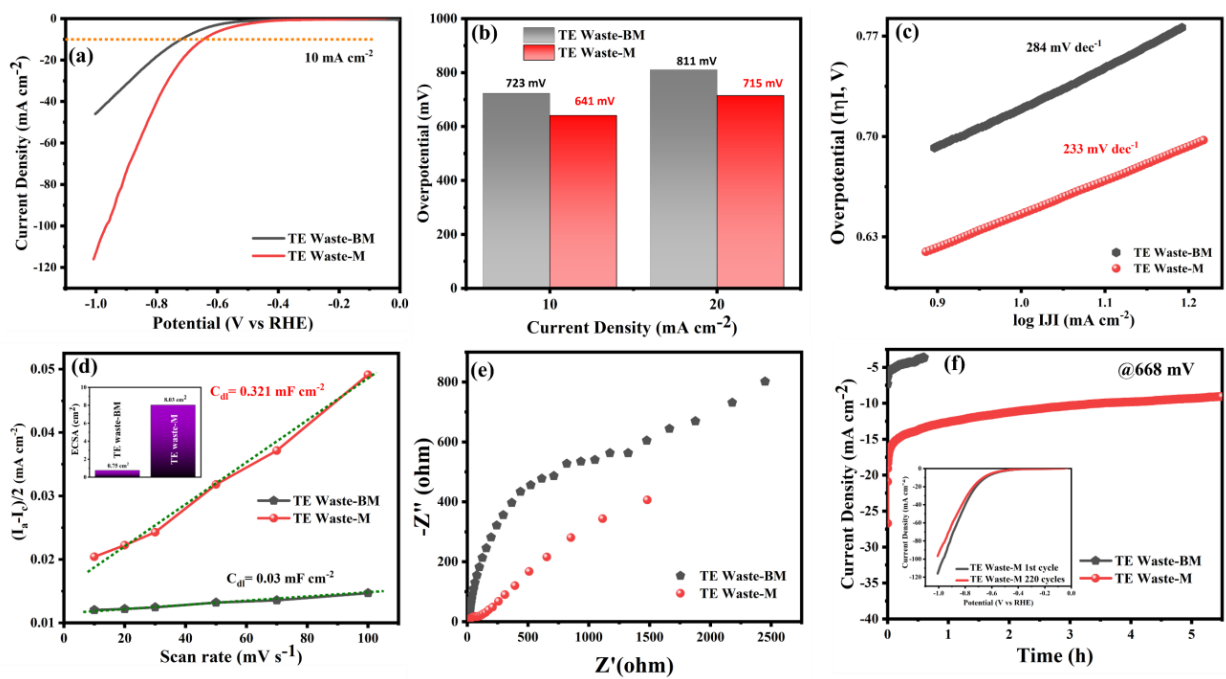

**Fig. 4** HER performance of M and BM catalysts in 1.0 M KOH electrolyte solution at a scan rate of 2 mV s$^{-1}$: (a) Curves of polarization (b) Comparison of overpotentials at current densities of 10 mA cm$^{-2}$; (c) The Tafel plots; (d) Plots illustrating the average capacitive current density vs scan rate (10, 20, 30, 50, 70, and 100 mV s$^{-1}$). The double-layer capacitance (Cdl) is equal to the linear slope; (e) Nyquist plots; (f) Thermoelectric E-waste of materials ball

milled and melted free-standing catalyst LSV curves before and after 300 cycles, as well as chronoamperometric observations of the HER at an overpotential of 159 mV.

To further understand the microstructural changes, SEM analysis before and after the HER has been performed for different samples and shown in **Fig. 5**. As seen in **Fig. 5a**, the low magnification image of the TE waste-BM sample before HER, showing agglomerated irregular powder particles. The alloy consists of three contrasts as evident in the BSE micrograph in **Fig. 5b**, region I- rich in Zn, region II- rich in Bi and Te and region III- suggesting BST phase. Thus, a Zn rich phase and $Bi_2Te_3$-$Sb_2Te_3$ solid solution is present in the TE waste-BM sample before HER. After the HER measurement, **Fig. 5c** shows the microstructure which is comparatively less agglomerated and also there is significant reduction in the particle size. From **Fig. 5d** it can be seen that there are two contrasts in the TE waste-BM: region I and II. Both regions show significant amount of oxygen and fluorine on the surface suggesting the surface oxidation or hydroxylation and fluoride deposition (possibly from electrolyte KOH, KF). In addition, there is no Zn observed in the sample after HER, suggesting the possible Zn dissolution during the electrochemical reaction. On changing the processing from ball milling to melting, TE waste-M sample shows comparatively smaller and more distributed particles (shown in **Fig. 5e**) before HER. The higher distribution and smaller particles is beneficial to expose more catalytic sites, enhance charge-transfer efficiency and offer a higher density of defect sites which assist fast hydrogen adsorption and evolution [34,35]. On further magnification, two different contrasts are observed in the sample: region -I and -II. Region -I shows the composition corresponds to BST while region -II suggests the presence of Zn based phase primarily ZnTe along with BST. After HER, TE waste-M sample shows relatively more porous and smoother morphology attributed to HER (**Fig. 5g**). In addition, there is high oxygen content suggesting the oxidation and hydroxylation of the sample during

HER. The region-I shows needle like morphology of the Zn-rich phase while the brighter phase corresponds to the BST which is being oxidized or hydroxylized during HER in alkaline electrolyte (shown in **Fig. 5h**).

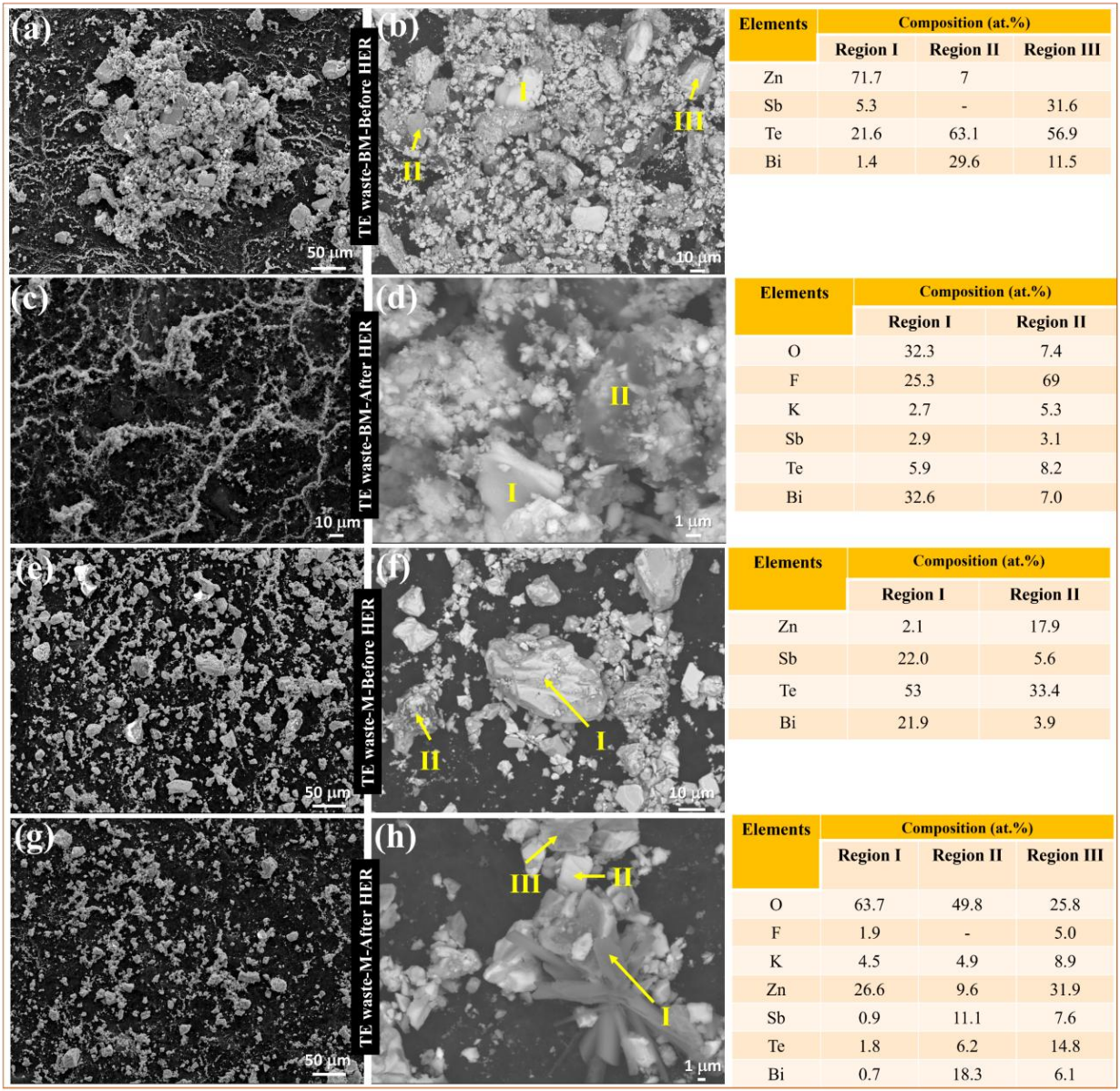

**Fig. 5** SEM morphology of TE-waste ball milled (BM) sample: (a)-(b) before and (c)-(d) after HER along with the elemental composition of the different contrasts. SEM morphology of TE-waste melted (M) sample: (e)-(f) before and (g)-(h) after HER.

XPS analyses after prolonged HER cycling reveals that the melted TE waste preserves its mixed oxidation states and defect-rich surface, with O 1s peaks corresponding to lattice oxygen, surface hydroxyls and adsorbed species. Notably, Te, Bi, Sb, and Zn largely maintain their partially oxidized states (**Fig. 6**), while, the ball-milled sample displays reduced Zn content, more amorphous features, and poorly defined oxidation states, indicative of a less homogeneous and less conductive surface (**Fig. 7**).[36,37] Although partial oxidation occurs in both samples during HER, the melted material reveals remarkable cyclic stability. Initially, a thin $TeO_x$ or Bi-O-Te layer is formed on the metallic/semi metallic sites, acting as a self-limiting passivation layer that shields the conductive core while permitting efficient electron transfer. The interface between the conductive Te-rich core and semiconducting oxide shell offers abundant oxygen-containing surface sites, enabling water adsorption and dissociation, crucial for the Volmer step of HER. In the ball-milled sample, lower Zn content and a more disordered structure led to fewer stable surface sites and a less robust oxide layer, resulting in lower cyclic stability. For the melted sample, oxide formation remains confined to the surface and dynamically equilibrates under HER potentials, evidenced by minimal changes in overpotential after 1000 CV cycles (if the oxidation were to extend deeper into the bulk, forming fully insulating oxides such as stoichiometric $TeO_2$ or $Bi_2O_3$, the electron transport pathway would be disrupted, leading to an increase in overpotential and degradation in activity).[38] Probably, in TE waste-BM after cyclic test, more insulating $TeO_2$ is formed as evident from **Fig. 7c**, which inhibit the electron transport path, while in melted TEW, some metallic Te still persists even after repetitive cyclic test (**Fig. 6c**), further assisting fast electron transport to enhance HER performance.[39] Moreover, the reversible surface redox equilibrium ensures that the thin $TeO_x$/Bi-O-Te layer functions both as a protective and catalytically active interface, enabling sustained hydrogen evolution. Overall, the higher homogeneity, mixed valence states, and retained surface hydroxyls in the melted sample account for its superior HER efficiency and long-term stability compared to the ball-milled counterpart.

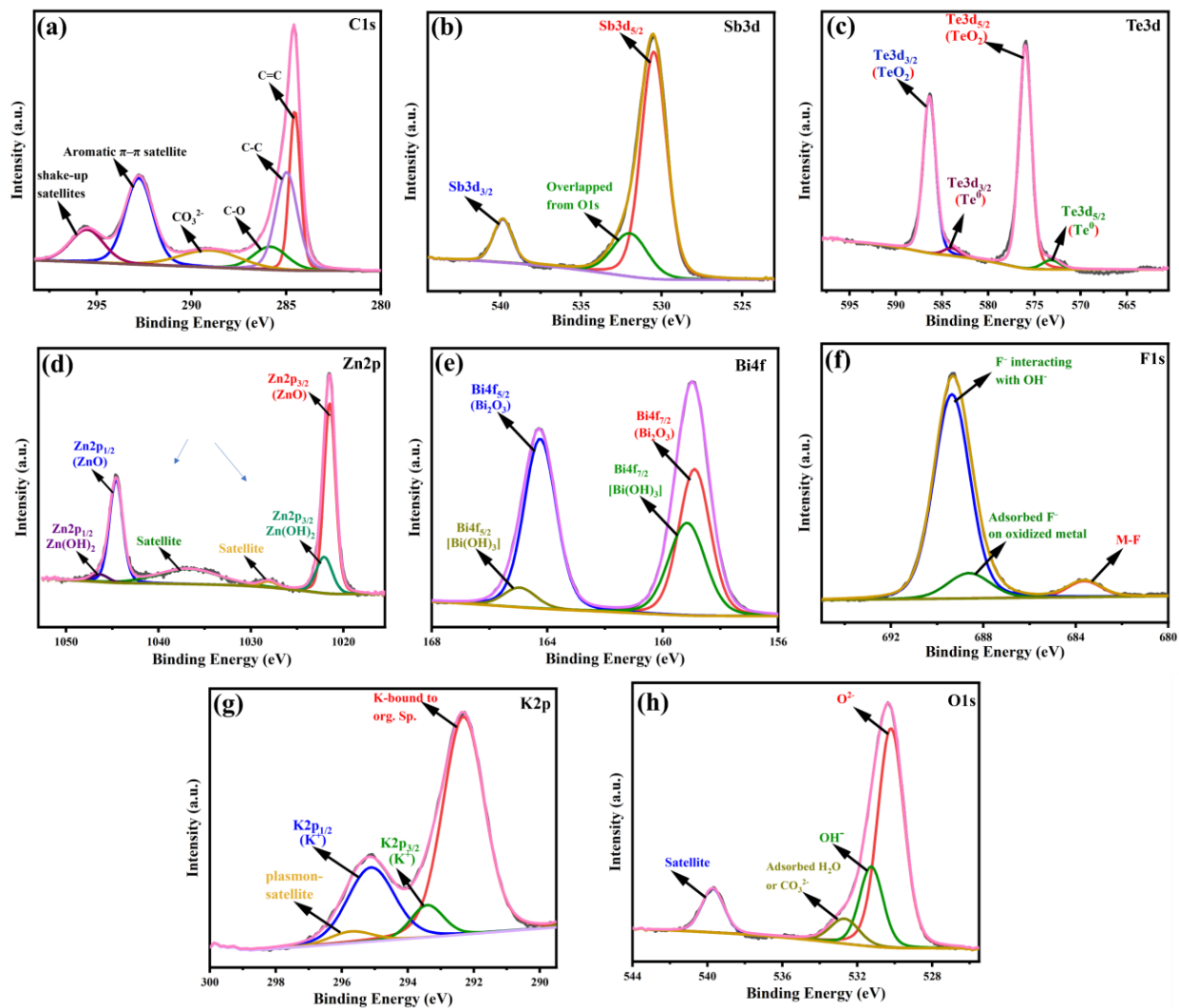

**Fig. 6** XPS deconvolution spectra for TE waste-M after HER cyclic stability: (a) C1s, (b) Sb3d, (c)Te3d, (d) Zn2p, (e) Bi4f, (f) F1s, (g) K2p, and (h) O1s.

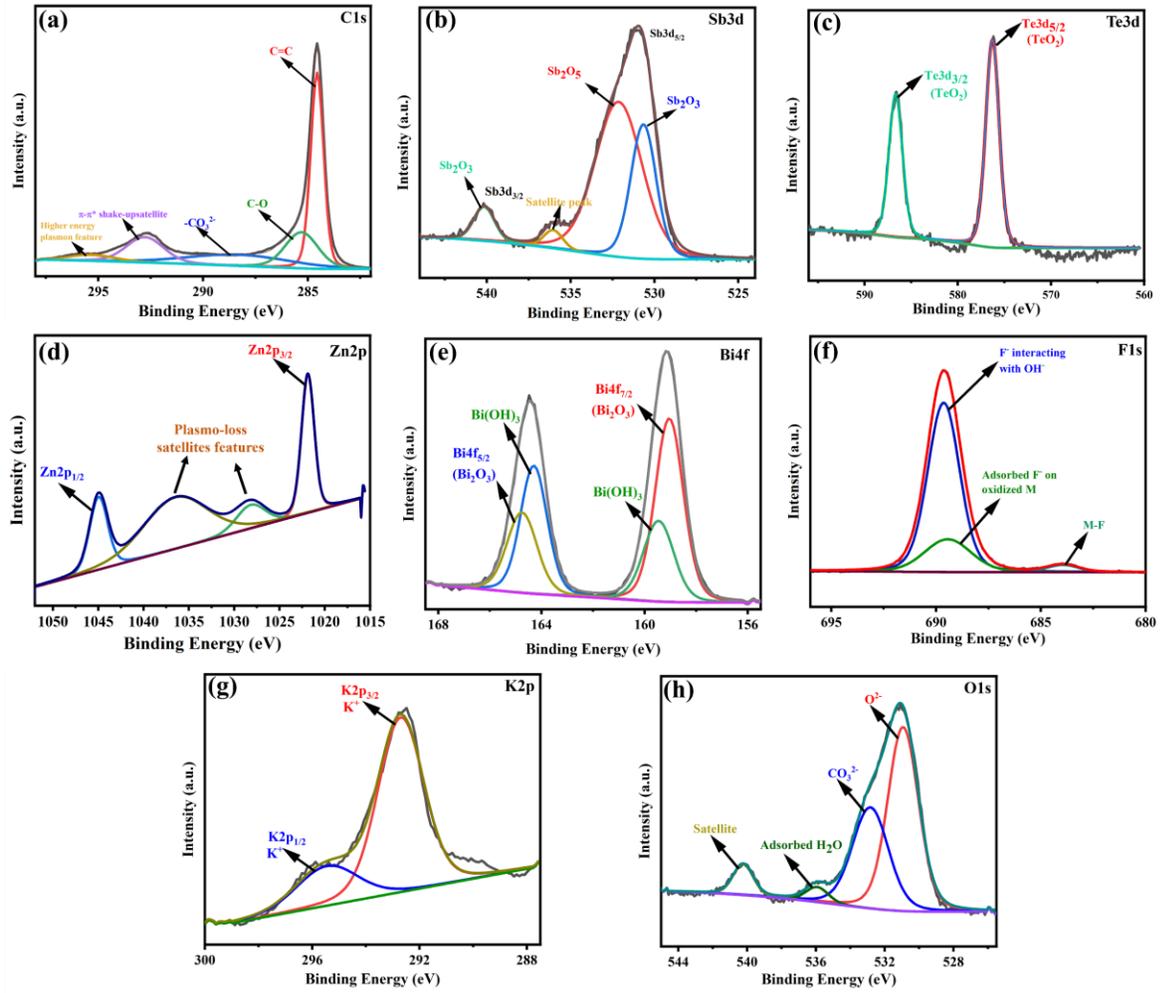

**Fig. 7** XPS deconvolution spectra for TE waste-BM after HER cyclic stability: (a) C1s, (b) Sb3d, (c)Te3d, (d) Zn2p, (e) Bi4f, (f) F1s, (g) K2p, and (h) O1s.

## 3.3. Theoretical Investigation of HER Performance of TE Waste

DFT calculations can provide atomistic insights on the interactions of the reaction intermediates with the structures, contributing to the understanding of the overall mechanism behind activity enhancement on the thermoelectric waste catalysts. Here, we have simulated different systems to investigate the effects of incorporating BST and ZnTe ($\alpha$) on the activity of $Bi_2Te_3$. First, we compare the bare $Bi_2Te_3$ with BST, simulated as a heterostructure of $Bi_2Te_3$ with $BiSbTe_3$. Then we investigate the activity of BST@$\alpha$ heterostructures.

To investigate the role of the number of layers on H adsorption, we have modelled systems with one and three monolayers. For a monolayer, we also consider both sides which present distinct terminations, as shown in **Fig. S2**. For each system, we consider different sites: top of the Bi and Te atoms as well as the bridge sites between them. Calculations were performed at unit cell size monolayer corresponding to full coverage for each site. The resulting adsorption free energies are shown in **Fig. 8a** where one can observe that pristine $Bi_2Te_3$ shows weak H adsorption, as seen by the large $\Delta G_H$ for all sites. Analyzing the role of the number of layers, it is interesting to observe that Top-Te and Bridge sites show less sensitivity to the number of layers, as seen by the small difference compared to the total value of the $\Delta G_H$. Also, Top-Te and Bridge sites showed the same activity for Monolayer-Bottom as well as the Top-Bi and Bridge sites for Monolayer Top. This result only reflects similar final configurations in these cases.

The $Bi_2Te_3$@BST heterostructures (shown in **Fig. 8b**) were modelled as $BiSbTe_3$ monolayer on top of $Bi_2Te_3$. The former was modelled considering several possible configurations where a half of the Bi atoms on the top $Bi_2Te_3$ monolayer are substituted by Sb. Next, the stability of each configuration was investigated calculating their formation energy ($E_{form}$). The results for $E_{form}$ are shown in supplementary **Fig. S3** and the most stable structure (index 50, shown in **Fig. 8b**) was used throughout the simulation. For the $\alpha$@BST heterostructures, a (111) ZnTe slab was constructed and assembled with the most stable BST structure, as shown in **Fig. 8b**. The electronic structure for $Bi_2Te_3$@BST is shown in **Fig. 8c** from which can be seen as a metallic behavior in contrast with the narrow semiconductor behavior of pristine $Bi_2Te_3$. Interestingly, two Dirac-like features are observed close to the M and K high symmetry points. Te p states dominate the valence band, whereas Bi, Sb and Te states hybridize at the conduction bands. For $\alpha$@BST heterostructure (electronic structure presented in **Fig. 8d**), the metallic behavior is also observed. In this case, the vicinity of Fermi level is less populated with states compared to the $Bi_2Te_3$@BST. On other hand, the valence states are also dominated by Te p state, while conduction bands present contributions from Bi, Sb, Zn and Te atoms.

For both $Bi_2Te_3$@BST and $\alpha$@BST, non-equivalent sites were selected for H adsorption. Also, in the case of the latter, both BST and ZnTe sides were considered for H adsorption. The selected sites for H adsorption on $Bi_2Te_3$@BST and $\alpha$@BST heterostructures and their corresponding H adsorption free energies are shown in **Figs. S4** and **S5** of supplementary information. The comparison of H adsorption free energies for all the systems is shown in **Fig. 8e**. $Bi_2Te_3$ has shown the worst H binding capacity among all the systems investigated, with a $\Delta G_H$ of 1.80 eV. This weak H binding capability agrees with previous studies [40], although caution should be taken when directly comparing the values due to differences in the exchange and correlation functional used by the authors. The $Bi_2Te_3$@BST ($\beta$@BST) heterostructures performed better compared to pristine $Bi_2Te_3$, exhibiting a $\Delta G_H$ of 0.15 eV. Next, we investigated $\alpha$@BST heterostructures probing the two sides: the BST edge and the $\alpha$ edge. The former performed worse compared to $\beta$@BST heterostructures with $\Delta G_H$ of 0.25 eV. The latter exhibited interesting results with two distinct sites: the first on top of Te with $\Delta G_H$ of –0.17 eV and the second close to Zn with $\Delta G_H$ of 0.60 eV. In terms of H binding capabilities, $\beta$@BST heterostructures have active sites for H binding on the ZnTe edge. To further investigate the origin of activity changes on each structure, we have calculated the Integrated Crystal Orbital Hamilton Population (ICOHP) for the best sites, which has been used as a descriptor for bond strength.[21,22] Here, we take the negative of the ICOHP to indicate that higher values as stronger binding. **Figure 8f** shows the -ICOHP for the best sites. It is possible to observe that the binding strength increases for the $Bi_2Te_3$@BST, $\alpha$@BST and BST@$\alpha$ heterostructures, compared to the $Bi_2Te_3$. Interestingly, $Bi_2Te_3$@BST and $\alpha$@BST heterostructures present close -ICOHP at the Top-Te sites. However, on the BST@$\alpha$ heterostructure, we observe the highest binding strength at the Top-Te site, in agreement with the negative H adsorption free energy. Therefore, incorporating BST and $\alpha$ phases on $Bi_2Te_3$ increases the formation of bonding states, thus increasing H binding strength ultimately reducing the H adsorption free energies.

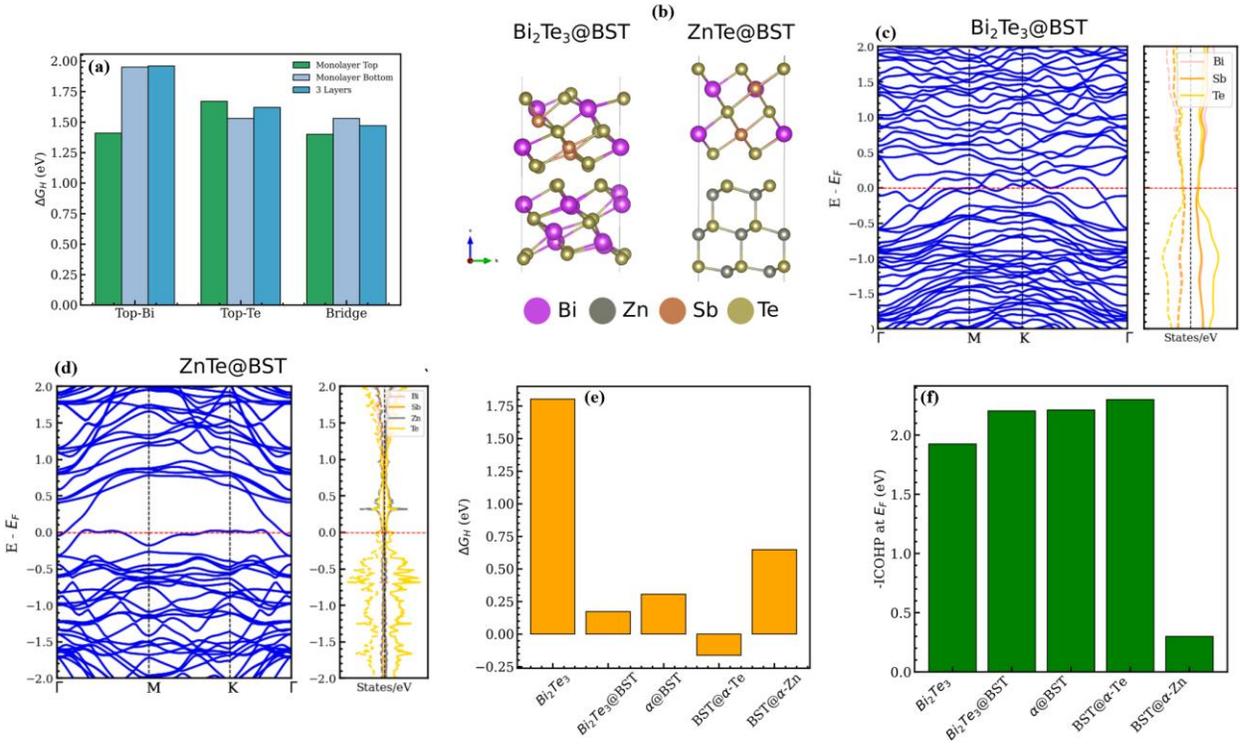

**Fig. 8** (a) H adsorption free energy as a function of the number of layers for pristine $Bi_2Te_3$. (b) Side view of the $Bi_2Te_3$@BST and ZnTe@BST heterostructures. electronic structure and projected density of states for $Bi_2Te_3$@BST heterostructure. (d) electronic structure and projected density of states of the ZnTe@BST heterostructure. (e) Comparison of H adsorption free energies on all the structures investigated. (f) Integrated Crystal Orbital Hamilton Population for all the structures.

## 4. Conclusion

The current work presents an effective circular economy and carbon-neutral approach for converting TEW into high-performance HER electrocatalysts in alkaline electrolyte. Processing of TEW by ball milling and melting paths developed multi component alloy with defect-enriched interfaces, tailorable phase compositions and preferable electronic structures, which in together promoted water dissociation and reduced hydrogen adsorption energy. Comparative HER studies reveal that TE Waste-M exhibited higher HER efficiency than TE Waste-BM-overpotential of 641 mV at 10 mA cm$^{-2}$, Tafel slope of 233 mV dec$^{-1}$, higher ECSA (8.03 cm$^2$) and long-term operational stability for 5.5 h without significant current loss. This may be due to the pronounced structural defect, formation of heterojunction of BiSbTe$_3$ and ZnTe, tunable electronic structure, grain boundary feature in TE Waste-M significantly expedite charge transfer and fast hydrogen-adsorption energy, beneficial for accelerating the efficiency for HER. First principles calculations were performed to investigate the activity of Bi$_2$Te$_3$, BST and ZnTe heterostructures, unveiling the effects of different phases and heterostructures on the H binding strength. Interestingly, incorporating BST and ZnTe increases the bonding states leading to increased ICOHP which ultimately shifts the H binding free energy towards zero, indicating increased activity. Moreover, the valorization of TE waste into functional HER catalyst eliminates the necessity for additional mining, complex chemical synthetic steps, precursor synthesis, thereby considerably decreasing environmental impacts and carbon footprint across the lifecycle. By coupling waste management resources with green hydrogen production, the present work offers an economically viable, scalable and low-carbon approach for designing HER catalyst. The outcome highlights the wider potential of TE waste-derived alloys in progressing sustainable energy technologies and lead to current global efforts towards accomplishing carbon-neutral hydrogen economies.

# References

1       R. Seif, F. Z. Salem and N. K. Allam, E-waste recycled materials as efficient catalysts for renewable energy technologies and better environmental sustainability, *Environ. Dev. Sustain.*, 2024, **26**, 5473–5508.

2       S. Mathi, H. Akhdar, R. S. Shetti, T. Alinad, A. N. Alodhayb, K. Mondal and N. P. Shetti, Amorphous electrocatalysts for oxygen and hydrogen evolution reactions: Advances in hydrogen production, *Mater. Today Sustain.*, 2025, **32**, 101223.

3       P. Bora, S. Kumar and D. Sinha, 2D transition metal dichalcogenides for efficient hydrogen generation, *Mater. Today Sustain.*, 2024, **27**, 100914.

4       S. Chattopadhyay, P. L. Mahapatra, M. N. Mattur, A. Pramanik, S. Gupta, T. S. Pieshkov, S. Saju, G. Costin, R. Vajtai, C. S. Tiwary, B. I. Yakobson and P. M. Ajayan, Unlocking the Potential: Atomically Thin 2D Fluoritene from Exfoliated Fluorite Ore and Its Electrochemical Activity, *Nano Lett.*, 2024, **24**, 7284–7292.

5       S. Chattopadhyay, C. C. de Oliveira, R. Bhar, D. Banik, T. S. Pieshkov, A. B. Puthirath, A. Pramanik, P. R. Sreeram, S. K. Saju, G. Costin, R. Vajtai, B. K. Dubey, K. Biswas, P. A. da Silva Autreto, C. S. Tiwary and P. M. Ajayan, Exploring Sustainable Hydrogen Production from Alkaline Fresh and Seawater Using Natural Ore Derived 2D $Bi_2S_3$, *Small*, DOI:10.1002/smll.202509283.

6       J. Torrubia, A. Valero and A. Valero, Energy and carbon footprint of metals through physical allocation. Implications for energy transition, *Resour. Conserv. Recycl.*, 2023, **199**, 107281.

7   É. Lèbre, M. Stringer, K. Svobodova, J. R. Owen, D. Kemp, C. Côte, A. Arratia-Solar and R. K. Valenta, The social and environmental complexities of extracting energy transition metals, *Nat. Commun.*, 2020, **11**, 4823.

8   I. Rey, V. L. Barrio and I. Agirre, Environmental assessment of a hydrogen supply chain using LOHC system with novel low-PGM catalysts: A life cycle approach, *Int. J. Hydrogen Energy*, 2025, **142**, 616–626.

9   W. Cai, S. Ito, E. Morioka, C. Chuaicham, A. Ulmaszoda, H. Miki and K. Sasaki, Synthesis of TiO2/copper-based oxide photocatalytic composites from copper smelting flotation slag for photocatalytic H2 evolution, *Mater. Today Sustain.*, 2025, **31**, 101215.

10  B. Sirichandana, R. Silviya, U. Sirimahachai, N. Patel and G. Hegde, Biowaste-derived porous nano-carbon-supported transition metal borides as efficient electrocatalysts for alkaline water splitting: a waste-to-wealth approach, *RSC Adv.*, 2025, **15**, 43275–43283.

11  R. Atchudan, S. Perumal, T. N. Jebakumar Immanuel Edison, S. Aldawood, R. Vinodh, A. K. Sundramoorthy, G. Ghodake and Y. R. Lee, Facile synthesis of novel molybdenum disulfide decorated banana peel porous carbon electrode for hydrogen evolution reaction, *Chemosphere*, 2022, **307**, 135712.

12  G. Kresse and J. Furthmüller, Efficient iterative schemes for ab initio total-energy calculations using a plane-wave basis set, *Phys. Rev. B*, 1996, **54**, 11169–11186.

13  S. Grimme, J. Antony, S. Ehrlich and H. Krieg, A consistent and accurate ab initio parametrization of density functional dispersion correction (DFT-D) for the 94 elements H-Pu, *J. Chem. Phys.*, DOI:10.1063/1.3382344.

14    P. E. Blöchl, Projector augmented-wave method, *Phys. Rev. B*, 1994, **50**, 17953–17979.

15    H. J. Monkhorst and J. D. Pack, Special points for Brillouin-zone integrations, *Phys. Rev. B*, 1976, **13**, 5188–5192.

16    J. K. Nørskov, J. Rossmeisl, A. Logadottir, L. Lindqvist, J. R. Kitchin, T. Bligaard and H. Jónsson, Origin of the Overpotential for Oxygen Reduction at a Fuel-Cell Cathode, *J. Phys. Chem. B*, 2004, **108**, 17886–17892.

17    J. K. Nørskov, T. Bligaard, A. Logadottir, J. R. Kitchin, J. G. Chen, S. Pandelov and U. Stimming, Trends in the Exchange Current for Hydrogen Evolution, *J. Electrochem. Soc.*, 2005, **152**, J23.

18    R. Dronskowski and P. E. Bloechl, Crystal orbital Hamilton populations (COHP): energy-resolved visualization of chemical bonding in solids based on density-functional calculations, *J. Phys. Chem.*, 1993, **97**, 8617–8624.

19    V. L. Deringer, A. L. Tchougréeff and R. Dronskowski, Crystal Orbital Hamilton Population (COHP) Analysis As Projected from Plane-Wave Basis Sets, *J. Phys. Chem. A*, 2011, **115**, 5461–5466.

20    I. Jamaï, N. Bekkioui, M. Ziati and H. Ez-Zahraouy, Effect of X (X = Sb, N, B) doping on structural, electronic, optical, photocatalytic, and thermoelectric properties of spinel $CdIn_2S_4$ for energy harvesting: A DFT approach, *Mater. Sci. Eng. B*, 2025, **313**, 117996.

21    R. Y. Rohling, I. C. Tranca, E. J. M. Hensen and E. A. Pidko, Correlations between Density-Based Bond Orders and Orbital-Based Bond Energies for Chemical Bonding Analysis, *J. Phys. Chem. C*, 2019, **123**, 2843–2854.

22  M. Khazaei, J. Wang, M. Estili, A. Ranjbar, S. Suehara, M. Arai, K. Esfarjani and S. Yunoki, Novel MAB phases and insights into their exfoliation into 2D MBenes, *Nanoscale*, 2019, **11**, 11305–11314.

23  F. Fei, Z. Wei, Q. Wang, P. Lu, S. Wang, Y. Qin, D. Pan, B. Zhao, X. Wang, J. Sun, X. Wang, P. Wang, J. Wan, J. Zhou, M. Han, F. Song, B. Wang and G. Wang, Solvothermal Synthesis of Lateral Heterojunction $Sb_2Te_3$/$Bi_2Te_3$ Nanoplates, *Nano Lett.*, 2015, **15**, 5905–5911.

24  Y. Du, B. Zhang, W. Zhang, H. Jin, J. Qin, J. Wan, J. Zhang and G. Chen, Interfacial engineering of $Bi_2Te_3$/$Sb_2Te_3$ heterojunction enables high–energy cathode for aluminum batteries, *Energy Storage Mater.*, 2021, **38**, 231–240.

25  L. Qi, W. Gao and Q. Jiang, Strain engineering of the electronic and transport properties of monolayer tellurenyne, *Phys. Chem. Chem. Phys.*, 2019, **21**, 23119–23128.

26  Y. Kim, H. Noh, B. D. Paulsen, J. Kim, I. Jo, H. Ahn, J. Rivnay and M. Yoon, Strain-Engineering Induced Anisotropic Crystallite Orientation and Maximized Carrier Mobility for High-Performance Microfiber-Based Organic Bioelectronic Devices, *Adv. Mater.*, DOI:10.1002/adma.202007550.

27  X. Yang, A. Banerjee, Z. Xu, Z. Wang and R. Ahuja, Interfacial aspect of $ZnTe$/$In_2Te_3$ heterostructures as an efficient catalyst for the hydrogen evolution reaction, *J. Mater. Chem. A*, 2019, **7**, 27441–27449.

28  X. Gao, N. Li, P. Li, Y. Wei, Q. Huang, K. Akhtar, E. M. Bakhsh, S. B. Khan, Y. Shen and M. Wang, $ZnTe$/$SnS_2$ heterojunction for photo-electrocatalysis of $CO_2$ to CO,

*Electrochim. Acta*, 2024, **497**, 144603.

29   S. Zhao, J. Xu, C. Wang, H. Zhang, Q. Liu and X. Kong, Enhancing edge chemistry in HCP-Ru catalysts through crystalline domain engineering for efficient alkaline hydrogen evolution, *Inorg. Chem. Front.*, DOI:10.1039/D5QI01143D.

30   L. Schultz, Formation of amorphous metals by mechanical alloying, *Mater. Sci. Eng.*, 1988, **97**, 15–23.

31   O. Norimasa, T. Kurokawa, R. Eguchi and M. Takashiri, Evaluation of Thermoelectric Performance of Bi2Te3 Films as a Function of Temperature Increase Rate during Heat Treatment, *Coatings*, 2021, **11**, 38.

32   Q. Hua, X. Chen, J. Chen, N. M. Alghoraibi, Y. Lee, T. J. Woods, R. T. Haasch, S. C. Zimmerman and A. A. Gewirth, Inducing Microstrain in Electrodeposited Pt through Polymer Addition for Highly Active Oxygen Reduction Catalysis, *ACS Catal.*, 2024, **14**, 7526–7535.

33   E. Westsson, S. Picken and G. Koper, The effect of lattice strain on catalytic activity, *Chem. Commun.*, 2019, **55**, 1338–1341.

34   Q. Gong, Y. Wang, Q. Hu, J. Zhou, R. Feng, P. N. Duchesne, P. Zhang, F. Chen, N. Han, Y. Li, C. Jin, Y. Li and S.-T. Lee, Ultrasmall and phase-pure W2C nanoparticles for efficient electrocatalytic and photoelectrochemical hydrogen evolution, *Nat. Commun.*, 2016, **7**, 13216.

35   Y. Dong, J. Ying, Y. Xiao, J. Chen and X. Yang, Highly Dispersed Pt Nanoparticles Embedded in N-Doped Porous Carbon for Efficient Hydrogen Evolution, *Chem. – An*

*Asian J.*, 2021, **16**, 1878–1881.

36   H. Y. Lee, I. J. Yang, J.-H. Yoon, S.-H. Jin, S. Kim and P. K. Song, Thermoelectric Properties of Zinc-Doped Indium Tin Oxide Thin Films Prepared Using the Magnetron Co-Sputtering Method, *Coatings*, 2019, **9**, 788.

37   A. Jilani, J. Iqbal, M. S. Abdel-Wahab, Y. Jamil and A. A. Al-Ghamdi, X-ray photoelectron spectroscopic (XPS) investigation of interface diffusion of ZnO/Cu/ZnO multilayer, *J. Optoelectron. Biomed. Mater.*, 2016, **8**, 27–31.

38   M. A. Qadeer, X. Zhang, M. A. Farid, M. Tanveer, Y. Yan, S. Du, Z.-F. Huang, M. Tahir and J.-J. Zou, A review on fundamentals for designing hydrogen evolution electrocatalyst, *J. Power Sources*, 2024, **613**, 234856.

39   Y. Liu, W. Wu and W. A. Goddard, Tellurium: Fast Electrical and Atomic Transport along the Weak Interaction Direction, *J. Am. Chem. Soc.*, 2018, **140**, 550–553.

40   Q. Qu, B. Liu, J. Liang, H. Li, J. Wang, D. Pan and I. K. Sou, Expediting Hydrogen Evolution through Topological Surface States on Bi2Te3, *ACS Catal.*, 2020, **10**, 2656–2666.